# Injection locking at 2f of spin torque oscillators under influence of thermal noise

M. Tortarolo[1*], B. Lacoste[1**], J. Hem[1], C. Dieudonné[1], M.-C. Cyrille[2], J. A. Katine[3], D. Mauri[3], A. Zeltser[3], L.D. Buda-Prejbeanu[1], U. Ebels[1]

[1]Univ. Grenoble Alpes, CEA, INAC-SPINTEC, CNRS, SPINTEC, 38000 Grenoble, France
[2]Univ. Grenoble Alpes, CEA-LETI MINATEC-CAMPUS, 38000 Grenoble, France
[3]HGST, 3403 Yerba Buena Road, San Jose, California 95135, USA

Experiments, numerical simulations and an analytic model were developed to elucidate the effects of noise in the synchronized state of a tunnel junction based spin torque nano oscillator (STNO). It is demonstrated that in the in-plane magnetized structure, while the frequency is locked, much higher reference currents are needed to reduce the noise by phase locking. Our analysis shows that it is possible to control the phase noise by the reference microwave current ($I_{RF}$) and that it can be further reduced by increasing the bias current ($I_{DC}$) of the oscillator, keeping the reference current in feasible limits for applications.

## I. INTRODUCTION

A spin polarized current passing through a magnetic multi-layered nanosystem can drive its magnetization into large amplitude periodic oscillations [1,2,3] when the spin polarized current is large enough to compensate the natural damping. These spin transfer driven magnetization oscillations, together with their particular nonlinear properties[4] spurred the interest in STNO's for several applications in current controlled microwave devices[5]. Nevertheless, one of the main issues that remains to be addressed for these spin STNO's is their relative large linewidth. One possibility to reduce the linewidth is to couple either different layers within an oscillator[6], or to couple several oscillators. For this second case, several options were proposed, experimentally and theoretically: current mediated coupling[7,8], dipolar coupling[9,10] or spin wave coupled nanocontacts[11,12]. In order to understand the conditions for electric synchronization of several oscillators by their own emitted RF current, we studied the synchronization of an STNO to a reference current source, with known spectral specifications. Here we focus on standard uniform in plane magnetized oscillators (in-plane polarizer and in-plane free layer, IP), for which an in-plane precession (IPP) mode is stabilized. The injection locking of such an STNO to a reference current at two times the generated frequency (2f) was demonstrated both numerically and by experiments[13]. However, the linewidth in the locked regime was reduced only by a factor of seven, while a reduction to the linewidth of the microwave source (several Hz) was expected. These large linewidths are associated to the thermal noise that induces fluctuations which can drive the phase from an equilibrium state to a neighbouring one, with an associated phase slip of ± 2π which can be envisaged as non-syncronization and re-synchronization events. Zhou *et al*[14] demonstrated that the particular way the phase approaches its synchronized value has consequences in the transients that may limit the modulation of an STO. Recent works investigated the mechanisms of the so called pure phase locking state in double vortex based STNO: Robust synchronization was experimentally shown, with a $10^5$ linewidth reduction[15] and the role of the phase slips in the synchronized state was investigated[16]. In this work we study the injection locking at *2f* to an external reference current of an uniform IP magnetized STNO under the influence of thermal noise. We performed both experiments and numerical simulations, together with an analytic model to describe the transients to the locked regime in the IPP geometry. Our results show the key features of electric synchronization of a uniform in plane magnetized STNO.

## II. ANALYTIC MODEL

The effect of thermal fluctuations on the transient behaviour of the synchronized state of an STNO is analyzed in the frame of a generic model of a non-linear auto oscillator[4] that we extended for the IPP mode synchronized by an RF current at *2f* (details in Appendix). Since STNO´s are non-linear (non-isochronous) oscillators, the power and the phase of the oscillator are not independent, leading to a system of coupled equations.

$$\frac{d\psi}{dt} = -\Delta\omega + 2N\delta p + \frac{1}{\sqrt{p_o}}\mathrm{Im}[\xi] \quad (1)$$

$$\frac{d\delta p}{dt} \cong -2\Gamma_p \delta p + 2p_0 \mathcal{F}\cos(\psi) + 2\sqrt{p_o}\,\mathrm{Re}[\xi] \quad (2)$$

Here $\psi(t) = 2\Phi - \omega_{ext}\cdot t$ is the phase difference between the STNO phase $\Phi$ and the phase of external source $\omega_{ext}\cdot t$, $N$ is the coefficient of non-linear frequency shift, $\mathcal{F}$ is a real parameter proportional to the reference current, $\Gamma_p$ is the damping rate of the power fluctuations and $\xi$ has the statistical properties of the Gaussian thermal noise[17]. Linearizing the equations (1) and (2) around a stable solution $p_o$ (without considering thermal noise) allows us to study the transient behaviour of the synchronized state, and analytically calculate the decay rate and the power spectral density (PSD) of the phase fluctuations $S_{\delta\Phi}$ of the synchronized state. In the limit of strongly non-linear oscillator $|\nu| \gg 1$ there are two solutions for the decay rate $\lambda$:

$$\lambda_{1,2} = \Gamma_p\left[1 \pm \sqrt{1 - \frac{\varepsilon}{\varepsilon_c}}\right] \quad (3)$$

Here, $\varepsilon = I_{RF}/I_{DC}$ and $\varepsilon_c = \Gamma_p^2/(2Np_0^2 \sin\psi_s P_x \Gamma_J |\mathcal{B}|/\mathcal{A})$ (see Apendix for the definition of the parameters). When $\varepsilon > \varepsilon_c$, $\lambda$ is complex with a real part given by $\Gamma_p$, that is the decay rate to the phase locked state and an imaginary part that describes an oscillatory approach to the phase-locked state with a frequency given by:

$$\Omega = \Gamma_p\sqrt{\varepsilon/\varepsilon_c - 1} \quad (4)$$

This is in agreement with Zhou *et al*[14], where they found for out of plane (OP) magnetized STNO's that the phase approaches its locked state exponentially and oscillating above a certain critical reference current.

Before discussing in more detail the oscillatory transient, we first will provide an expression for the phase noise in the synchronized state. By taking into account the thermal noise, we can calculate from eq. 1,2 the power spectral density (or single sideband) $S_{\delta\Phi}$ of the phase fluctuations of the synchronized state:

$$S_{\delta\Phi} = 2\pi\Delta f_o \frac{\left(\frac{\Gamma_p}{\pi}\right)^2 (1+\nu^2) + f^2}{\left[-\left(\frac{\Gamma_p}{2\pi}\right)^2 \frac{\varepsilon}{\varepsilon_c} + f^2\right]^2 + 4\Gamma_p^2 f^2} \quad (5)$$

Here $\nu = Np_o/\Gamma_p$ is the dimensionless nonlinear frequency shift and $\Delta f_o$ is the free running linewidth. Eq. 5 is plotted in Fig. 1a with the parameters calculated from the analytical model (see Appendix) for the free running state with a bias current $I_{DC} = 50 \times 10^{10}$ A/m$^2$, which leads to an IPP stable precession mode around 4.7 GHz and a $\Delta f_o = 50$ MHz. In this configuration the system has a coefficient of nonlinear frequency shift $N = -3.16 \cdot 10^{10}$ rad/sec, a damping rate of the power fluctuations $\Gamma_p = 666$ Mrad/sec, a normalized dimensionless nonlinear frequency shift parameter $|\nu| = 16$, and $\varepsilon_c = 0.025$. Since this value of $\varepsilon_c$ for these uniform IP STNO's is small compared to $\varepsilon$ (~0.1 or higher) the locking to the synchronized state always takes place via an oscillatory transient.

The characterization of the phase noise properties by the PSD in the Fourier space has the advantage that its inverse power law dependence on frequency PSD $\sim 1/f^x$ provides information about underlying noise processes. The model predicts a crossover from $1/f^2$ to $1/f^0$ with increasing reference current, with the two limit cases:

$$S_{\delta\Phi} = \begin{cases} f \gg f_{roll-off} : \frac{2\pi\Delta f_o}{f^2}\left(\frac{1}{f^2 + 4\Gamma_p^2}\right) \\ f \ll f_{roll-off} : 2\pi\Delta f_0 \frac{4(1+\nu^2)}{(\Gamma_p/\pi)^2} \frac{\varepsilon_c}{\varepsilon} \end{cases} \quad (6)$$

As already shown experimentally[18] for IPP STNO's, the free running oscillator ($\varepsilon = 0$) shows a $1/f^2$ dependence associated to a random walk of the phase (blue line, Fig. 1a). This behaviour is modified when applying a reference at 2f: Even for a low external force ($\varepsilon = 0.1$, yellow line, Fig.1a) below the roll of frequency $f_{roll\ off} \sim 1/\Gamma_p$ down to the lowest (calculated) offset frequency the phase noise is constant. This corresponds to fluctuations of the phase around its locked value. The Eq. (6) shows that the phase noise level in this region can be decreased upon increasing the reference current $\varepsilon$. Above $\varepsilon = 0.1$ there is a peak around $f_{roll\ off}$ that is related with the oscillatory relaxation mechanism[14,20,21]. This will be discussed in more detail in the next section.

III. MACROSPIN ANALYSIS

We performed macrospin simulations for the in-plane precession (IPP) mode of an in-plane magnetized polarizer and free layer MTJ, using a solver for the Landau-Lifshitz-Gilbert equation and taking into account the damping like torque term (the field like term was neglected in this work, see Appendix). The simulation parameters are as follows: free layer of size 90x80x3.9 nm$^3$; spontaneous magnetization $M_s = 1000$ kA/m, damping parameter $\alpha = 0.02$ and zero magneto-crystalline anisotropy. The polarizer is aligned in the plane at 165° from the free layer magnetization equilibrium position and a spin-polarization $\eta = 0.37$ is supposed. A static magnetic field of 40 mT was applied along the in plane easy axis (Ox). The continuous current was set to $J_{DC}$ = -50·10$^{10}$ A/m$^2$, leading to an IPP stable precession mode with f ~ 4.7 GHz. A white Gaussian thermal noise field was added, corresponding to 10 K, 20 K, 50 K, and 100 K[13,21]. We present here the results for 50 K, similar behavior was obtained for all the studied temperatures. The frequency of the RF current was set to two times the free running STNO frequency $2f_0$=9,5396 GHZ, which corresponds to the centre of the locking range.

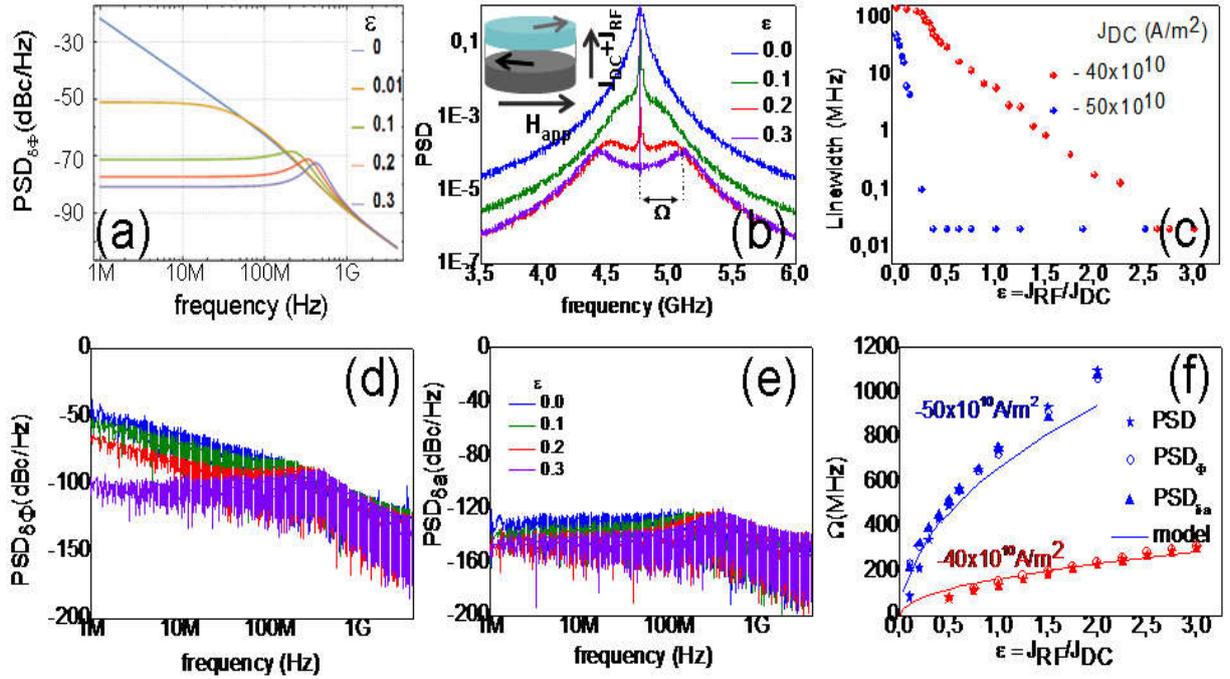

FIG.1. Phase noise from analytic (a) and numerical simulations at 50K (d). Notice the peaks appearing ~200 - 300 MHz. (b) The PSD of the signal as a function of ε and schematics of the oscillator (inset) for $J_{DC}$ = -50x10$^{10}$ A/m$^2$. (c) FWHM vs. ε for low current $J_{DC}$= -40x10$^{10}$ A/m$^2$ (red dots) and medium current $J_{DC}$ = -50x10$^{10}$ A/m$^2$ (blue dots). For ε> 2 (low $J_{DC}$ regime), the linewidth falls below the resolution of the technique.(f) Sideband frequency for both low current regime (red set) and medium current regime (blue set) extracted from the PSD (stars), and from the peaks on the phase noise(Fig. 1d) (open circle). The solid line corresponds to the analytical model developed for the IPP geometry (eq. 4).

The phase and amplitude noise in the synchronized state as a function of the reference current was extracted from the simulated temporal traces of the $m_y$ component of the magnetization (in-plane magnetization along the short axis of the pillar) using the Hilbert transform method[22,23] which allows the reconstruction of an analytic signal from the voltage output:

$$V = V_0 (1+\delta a)\cos(2\pi ft + \varphi) \quad (7)$$

The $S_{\delta\Phi}$ at 50K extracted from the numerical time integration of the LLG equation is shown in Fig. 1d. The corresponding evolution of the power spectral density of the $m_y$ component of the magnetization with ε and its FWHM are displayed in Fig. 1b and 1c respectively. Both phase and amplitude noise (Fig. 1d and e) decrease with the reference current and a clear crossover from a $1/f^2$ to a $1/f^0$ (white noise) is seen upon increasing it.

Before addressing the phase noise level in comparison to the analytic results and the linewidth, we now discuss the PSD of the $m_y$ component of the magnetization, Fig. 1b. The peak of the free running state becomes very narrow and two symmetric sidebands appear upon increasing the reference current (Fig.1b). These sidebands are also visible on $S_{\delta\Phi}$, that shows a peak around $f_{roll-off}$ whose frequency depends on ε. All these features are consistent with the analytic model, where $S_{\delta\Phi}$ calculated with the parameters listed above (eq.5, Fig. 1a) shows the peaks associated to the sideband frequencies and predicts the oscillatory transient with frequency given by eq. (4). Figure 1f shows the frequency of these sidebands extracted from the PSD of the $m_y$ component of the magnetization and from the numeric $S_{\delta\Phi}$ for two different bias currents $I_{DC}$= -40x10$^{10}$ A/m$^2$ (red symbols) and $I_{DC}$= -50x10$^{10}$ A/m$^2$ (blue symbols). The full line represents the model (eq.4) for both bias currents. This comparison confirms that the peaks of the phase noise and the sidebands have the same physical origin arising from the oscillatory approach of the transient. Furthermore the comparison supports the analytic model.

In the following we discuss the phase noise level of the numerical results. There are two contributions to the phase noise in the synchronised state. The first one, as was discussed for the analytical description are phase fluctuations around the stable phase that is given by the external source plus a constant phase shift. The second contribution are phase slips[24], not considered in the analytical model but that are present in the numerical calculation. To understand their contribution to the phase noise and linewidth we extracted the phase from simulated time traces for different reference current values (Fig. 2c). The phase trace shows a drift in time, together with the appearance of the phase slips, which become well defined upon increasing the reference current (Fig.2a). As can be seen the number of phase slips per sampling period (~40 μs) decreases. These phase slips are responsible for the $1/f^2$ contribution of the phase noise at low offset frequencies. To demonstrate this, we compare the phase noise extracted from different 10 μs segments of the total 40 μs phase trace that contain respectively none, one or two phase slips. In Fig. 2b it is clearly seen that in presence of phase slips the phase noise has a $1/f^2$ dependence at low frequency f <$f_{roll\ off}$, while in absence of phase slips the phase noise level is constant. The increase of the phase noise level due to phase slips is also expressed in an enhanced linewidth, see Fig. 1c.

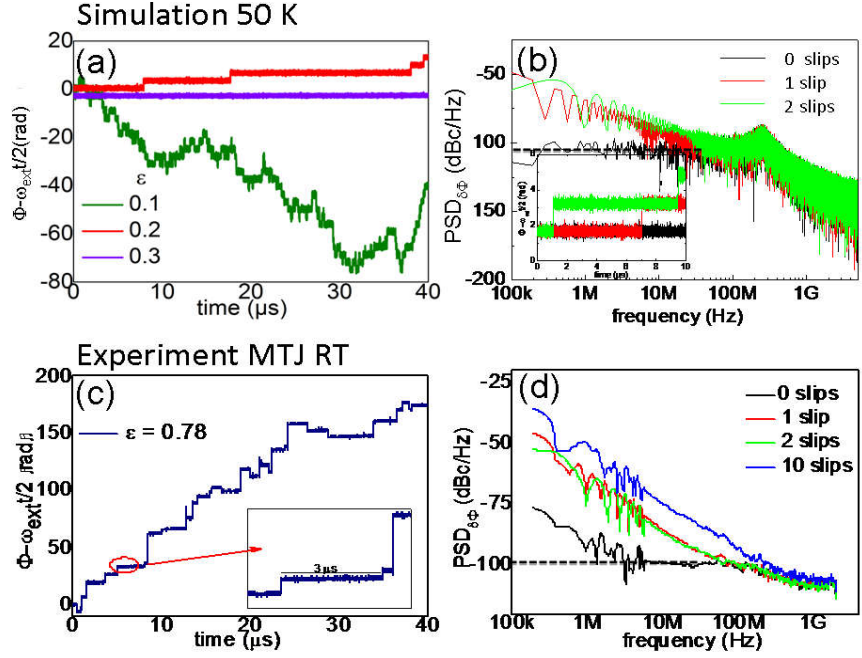

FIG.2. Simulated (a) and experimental (c) phase temporal traces. Inset: detail of 5μs segments of the temporal trace. The phase slips decrease in number with increasing ε disappearing at ε = 2.75. (b) Phase noise analysis on the temporal trace segments (inset) corresponding to no phase slips (black), 1 phase slip (red) and 2 phase slips (green) . (b) Phase noise analysis from the experimental time trace from 3s segments with 0, 1,2,10 phase slips.

From the numerical analysis we can see that the drastic decrease in the linewidth with ε (Fig. 1c) can be related to the decreasing number of phase slips. Particularly, when the phase noise flattens for ε>0.3 (ε> 2.75) for $I_{DC}$= -50x10$^{10}$ A/m$^2$ ($I_{DC}$= -40x10$^{10}$ A/m$^2$), the phase slips are absent in the phase trace and the linewidth falls under the resolution limit of the numerical calculation (20 kHz). This fact raises the question whether the so called "pure" synchronization[15,16] is due the absence of phase slips, where the STNO would reduce its linewidth ideally to the one of the reference source. In the case discussed here , this means that for values larger than ε = 0.3 (ε = 2.75) torques from the reference current on the magnetization are strong enough to stabilize the phase around a single value and the remaining noise is given by the one discussed within the analytical model, describing damped oscillations around the stable phase, for noise frequencies f <$f_{roll\ off}$ . We point out here that the absence of the phase slips depends on the observation time, i.e. the length of the temporal trace: longer observation times increase the probability of phase slips. These results evidence that even if the system is in the frequency locking regime, higher values of reference current are needed to achieve full linewidth reduction by phase locking. This observation is in agreement with Lebrun *et al*[16] that highlighted the difference between the reported "frequency locked state"[13,26,27] and pure phase locked states in absence of phase slips for vortex oscillators with a free running frequency of ~200 MHz and ~100 kHz linewidth.

IV. EXPERIMENT

The analytical and numerical results explain qualitatively the experimental observations on the synchronisation by a reference current. The studies were realized on the same type of devices presented in Ref.[13,18,28], which are in plane magnetized magnetic tunnel junctions (MTJ), having a stack composition of IrMn/CoFeB/Ru/CoFeB/MgO/CoFe/CoFeB and nominal resistance area RA=1 $\Omega\mu m^2$. While results were obtained on different devices, we present here results on for one device. Its autonomous, i.e. free running regime is characterized by a free running frequency of $f_0$= 7.5 GHz for a bias current $I_{DC}$= -1.6 mA and an applied in plane field of 350 Oe, with a linewidth of 55 MHz. The synchronization experiment was done varying the reference current frequency around two times the free running frequency (2f) of the oscillator, from 14 GHz to 16 GHz, and the source power was varied from -15 dBm to -5 dBm (corresponding to a reference current of ~0.3 to 1.3 mA), just before the sample starts to show signs of degradation. A detailed description of the experiment is available in Ref. [13]. The temporal traces were measured using a single shot oscilloscope[13,29,30], and amplitude and phase noise were extracted using the same protocol as for the simulated data. The PSD map of the output powerfor the STNO frequency f as a function of the source frequency $f_{ext}$ is shown in the Fig. 3a for $\varepsilon=J_{RF}/J_{DC}$=0.7. In Figure 3b it is clearly seen that for increasing reference current $\varepsilon$, the frequency locking range is wider, and that the linewidth decreases until its minimal value. Notice that the linewidth reaches a 10x reduction (8 MHz with a 1 MHz resolution bandwidth). The amplitude noise shows a $1/f^0$ behavior both for synchronized (Fig.4c, grey dashed line) and the free running state (Fig.3c, black full line), characteristic of white noise fluctuations of the amplitude around its stable value. The experimental plots of the phase noise in the synchronized (Fig. 4c, red dashed line) and the free running state (red full line) show that the synchronization mechanism is efficient to reduce the phase noise by 20 dBm with respect to the non-synchronized state. corresponding to the observed relatively large linewidth of ~ 8 MHz, instead of the expected complete reduction to the linewidth of the reference source (few Hz). Both plots display a $1/f^2$ behaviour characteristic. However the origin is different. In the free running state it results from a random walk of the phase, while in the synchronised state it is due to the phase slips as explained in section III. Our experiments show that for the maximum applied reference current $I_{rf}$, even though the oscillator is synchronized with the external source, the emission linewidth remains broad. The phase noise decreases but it does not reach the constant level at which the linewidth is expected to be reduced to the source noise. This was not observed in our experiments because the voltage breakdown of the samples did not allow to continue increasing the reference current preventing the STNO from achieving a pure phase locked state. For the same reason, we were not able to observe sidebands in the experimental voltage output. Nevertheless, we have extracted the phase noise from shorter 3 μs segments of the temporal trace[25] of the phase (Fig2.c), that include either 0, 1, 2 or 10 phase slips (Fig.2d). As can be seen in absence of phase slips the phase noise is flat in a certain range of offset frequencies ~5-100MHz in Fig. 2d. This demonstrates the lowest phase noise level that can be reached for the in-plane STNO, when phase slips would be completely suppressed. Note that this level is with -100dBc/Hz the same as in the simulations, Fig. 2b and also the same as has been reported for vortex devices[16].

The relative large linewidth of ~ 50 MHz of our STNO's could be a drawback to linewidth reduction as discussed by Hamadeth *et al*[15] where by decreasing 7 times the free running linewidth, the linewidth reduction in the locked regime goes from only 10 to $10^5$. Indeed, this is also witnessed in our numerical simulations, where the smaller free running linewidth at higher bias current ($J_{DC}$= -50x$10^{10}$A/m$^2$) leads to a linewidth reduction for significantly lower reference currents.

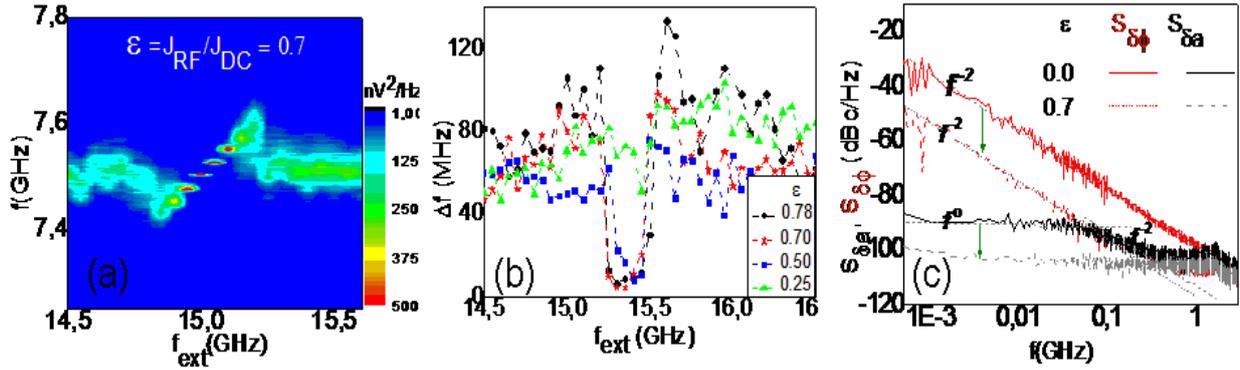

FIG.3: PSD map of the output voltage at $I_{RF}$ = 1.12 mA (a). Linewidth vs $\varepsilon$= $I_{RF}/I_{DC}$(b) and amplitude and phase noise from the experiment (c) for the non-synchronized state, (continuous line, $\varepsilon$=0) and synchronized state (dashed line, $\varepsilon$=0.7). Notice that the synchronization mechanism is efficient to reduce the noise level (green arrows).

## V. CONCLUSION

We have studied the synchronization mechanism of an uniform IP magnetized STNO under thermal noise. The synchronization of these devices was demonstrated in several experiments, however no more than a 10 times reduction of linewidth was achieved. This is explained by numerical simulations including thermal noise. While the STNO can be synchronized by moderate rf currents, higher rf currents are needed for full linewidth reduction. Full linewidth reduction is achieved when phase slips are suppressed. With increasing reference current the number of phase slips is reduced resulting in a crossover from $1/f^2$ to $1/f^0$ behaviour in the phase noise when the phase slips are suppressed. Experiment and simulations indicate that the lowest phase noise level achievable under synchronisation is on the order of -100 dBc. The simulations also shows that it is possible to achieve linewidth reduction for lower reference currents by increasing the bias current of the oscillator. This study will be important for designing STNO configurations of appropriate performances for microwave applications in the gigahertz range.

## ACKNOWLEDGMENT

This work was supported in part by the French National Research Agency (ANR) under contract N° 2011 Nano 01607 (SPINNOVA) and in part by the EC under the FP7 program N° 317950 MOSAIC.

\* M.T. is now at CONICET and Universidad Nacional de San Martin, Buenos Aires, Argentina.

\*\*B. L. is at International Iberian Nanotechnology Laboratory, Braga, Portugal

APPENDIX A: Model for 2f synchronization in IPP mode

The transient behavior in the synchronized state of a STNO is analyzed in the frame of a generic model of a nonlinear auto-oscillator. The model proposed by ref. [4] is extended for the synchronization by an RF spin-current at 2f. The configuration selected here has both the free-layer and the polarizer in-plane magnetized. The magnetization of the free layer is supposed to be uniform, thus the Gibbs free energy associated of the ferromagnetic free layer of the nanopillar is:

$$E(\mathbf{m}) = K_u(1-m_x^2)V - \mu_0 M_s H_0 m_x V + \frac{1}{2}\mu_0 M_s^2 (N_X m_x^2 + N_Y m_y^2 + N_Z m_z^2)V \quad (A1)$$

Where $\mathbf{m}=\mathbf{M}/M_s$ is unitary magnetization vector, V the volume of the sample, $M_s$ the spontaneous magnetization. A static magnetic field of amplitude $H_0$ is applied in the plane of the sample along the Ox direction. The demagnetizing effects are accounted by the demagnetizing tensor $\mathbf{N}=(N_X, N_Y, N_Z)$ and a uniaxial magneto-crystalline anisotropy along Ox direction is considered of amplitude $K_u$ ($K_u>0$). Following the Holstein-Primakoff transformation the variables $m_x$, $m_y$, $m_z$ are replaced by the canonical variables (a, a\*) such as:

$$a = \frac{m_y - jm_z}{\sqrt{2(1+m_x)}} \quad (A2)$$

It is convenient to express the reduced Gibbs free energy in reduced units:

$$\frac{\gamma_0'}{\mu_0 M_s V} E(a,a^*) = \mathcal{A}aa^* + \frac{1}{2}\mathcal{B}(a^2 + a^{*2}) + \mathcal{V}(aa^{*3} + a^3 a^*) + \mathcal{U}a^2 a^{*2} \quad (A3)$$

Where $\gamma_0' = \frac{\mu_0 \gamma}{1+\alpha^2}$ with α is the damping constant, $\gamma_0 = \mu_0 \gamma$ is the gyromagnetic ratio of the free electron multiplied by the vacuum permeability $\mu_0$. The notations are similar to that of Ref. 4.

$$\begin{aligned}
\mathcal{A} &= \omega_A + \omega_H + \omega_M/2 \\
\mathcal{B} &= -\omega_M/2 \\
\mathcal{U} &= -(\omega_A + \omega_M/2) \\
\mathcal{V} &= \omega_M/4 \\
\omega_M &= \gamma_0' M_s (N_Z - N_Y) \\
\omega_H &= \gamma_0' H_0 \\
\omega_A &= \gamma_0' \left[\frac{2K_u}{\mu_0 M_s} + M_s(N_Y - N_X)\right]
\end{aligned} \quad (A4)$$

A second transform is used for the diagonalization of the quadratic part of the reduced Gibbs free energy (Hamiltonian): $b=ua+va^*$ where

$$u = \sqrt{\frac{\mathcal{A}+\omega_0}{2\omega_0}}, v = -\sqrt{\frac{\mathcal{A}-\omega_0}{2\omega_0}} \quad (A5)$$

with $\omega_0 = \sqrt{\mathcal{A}^2 - \mathcal{B}^2}$. The last transformation is simply a normalization: $b = \sqrt{\frac{\omega_0}{\mathcal{A}}}c$.

Once a spin-polarized current of polarization **p**=($P_x$,$P_y$, 0) is injected in the sample the modified Gilbert equation is used to describe the magnetization dynamics considering the damping-like term of the spin-transfer:

$$\frac{d\mathbf{m}}{dt} = -\gamma_0(\mathbf{m}\times\mathbf{H}_{eff}) + \alpha\left(\mathbf{m}\times\frac{d\mathbf{m}}{dt}\right) - \gamma_0 a_J J_{app}\mathbf{m}\times(\mathbf{m}\times\mathbf{p})$$
(A6)

The effective field is given by the functional derivative of the Gibbs free energy with respect to the magnetization: $\mathbf{H}_{eff} = -\frac{1}{\mu_0 M_s V}\frac{\delta E}{\delta \mathbf{m}}$. The injected current density is time dependent and given by:

$$J_{app}(t) = J_{DC} + J_{RF}\cos(\omega_{ext}t) = J_{DC}[1+\varepsilon\cos(\omega_{ext}t)]$$

The spin–torque amplitude coefficient is $a_J = \frac{\hbar}{2e}\frac{\eta}{\mu_0 M_s t}$ where $t$ is the thickness of the free layer and $\eta$ is the spin-polarization.

The numerical analysis is carried out on the equivalent modified Landau-Lifshitz equation:

$$\frac{d\mathbf{m}}{dt} = -\gamma_0'(\mathbf{m}\times\mathbf{H}_{eff}) - \alpha\gamma_0'[\mathbf{m}\times(\mathbf{m}\times\mathbf{H}_{eff})]$$
$$-\gamma_0' a_J J_{app}\mathbf{m}\times(\mathbf{m}\times\mathbf{p}) + \gamma_0'\alpha a_J J_{app}(\mathbf{m}\times\mathbf{p})$$
(A7)

Applying the three transformations presented above $\mathbf{m}\rightarrow(a,a^*)\rightarrow(b,b^*)\rightarrow(c,c^*)$, keeping only the potential resonant terms the following equation for the complex variable $c$ is obtained:

$$\frac{dc}{dt} = -j[\omega_0 + N|c|^2]c - \Gamma_0[1+Q_1|c|^2+Q_2|c|^4]c$$
$$-\Gamma_J P_x[1+\varepsilon\cos(\omega_e t)]\left[1-|c|^2-\frac{|\mathcal{B}|}{2\mathcal{A}}(c^2+c^{*2})\right]c$$
$$+\Gamma_J P_y[1+\varepsilon\cos(\omega_e t)]\frac{1}{2}\sqrt{\frac{|\mathcal{B}|}{\omega_0}}(u+v)$$
(A8)

The expressions for the coefficients are the following:

$$N = -[3\omega_M uv(u^2+v^2)+(2\omega_A+\omega_M)(u^4+v^4+4u^2v^2)]\frac{\omega_0}{\mathcal{A}}$$
$$\Gamma_0 = \alpha\mathcal{A}$$
$$Q_1 = -\left[3uv\omega_M + (u^2+v^2)\left(\omega_H+3\omega_A+\frac{3}{2}\omega_M\right)\right]\frac{\omega_0}{\mathcal{A}^2}$$
$$Q_2 = [3uv(u^2+v^2)\omega_M + (u^4+4u^2v^2+v^4)(2\omega_A+\omega_M)]\left(\frac{\omega_0}{\mathcal{A}}\right)^2$$
$$\Gamma_J = \gamma_0' a_J J_{DC}$$
(A9)

In Eq. A8 it is important to keep the term involving $c^2+c^{*2}$ since it mediates the 2f synchronization. This has been neglected until now in the literature.

The analysis is continued in terms of power $p$ and phase $\Phi$ such as $c=\sqrt{p}e^{-j\Phi}$ for which two coupled equations are obtained:

$$\begin{cases}\frac{dp}{dt} = -2\Gamma_0[1+Q_1 p]p - 2P_x\Gamma_J[1+\varepsilon\cos(\omega_e t)]\left[1-p-\frac{|\mathcal{B}|}{\mathcal{A}}p\cos(2\Phi)\right]p \\ \qquad +2P_y\Gamma_J[1+\varepsilon\cos(\omega_e t)]\frac{1}{2}\sqrt{\frac{|\mathcal{B}|}{\omega_0}}(u+v)\sqrt{p}\cos\Phi \\ \frac{d\Phi}{dt} = [\omega_0+Np] - P_y\Gamma_J[1+\varepsilon\cos(\omega_e t)]\frac{1}{2}\sqrt{\frac{|\mathcal{B}|}{\omega_0}}(u+v)\frac{\sin\Phi}{\sqrt{p}}\end{cases}$$
(A10)

In DC regime ($J_{RF}$=0) these generic equations allow to extract the power $p_0$ in the free-running state:

$$p_0 = -\frac{\Gamma_0+P_x\Gamma_J}{\Gamma_0 Q_1-P_x\Gamma_J} \qquad (A11)$$

and to define the amplitude relaxation rate:

$$\Gamma_p = (\Gamma_0 Q_1 - P_x \Gamma_J) p_0. \qquad (A12)$$

In the 2f synchronized state the phase difference between the STNO and the external signal $\psi = 2\Phi - \omega_{ext} t$ is constant and the stationary power $p_s$ is shifted by $\delta p = p_s - p_0$. Keeping the resonant terms at 2f, it is possible to obtain the coupled equations:

$$\begin{cases} \dfrac{d\psi}{dt} = -\Delta\omega + 2N\delta p \\ \dfrac{d\delta p}{dt} \cong -2\Gamma_p \delta p + 2 p_0 \mathcal{F} \cos(\psi) \end{cases} \qquad (A13)$$

with $\Delta\omega = \omega_{ext} - 2(\omega_0 + N p_0)$ and $\mathcal{F} = \varepsilon P_x \Gamma_J \dfrac{|\mathcal{B}|}{2\mathcal{A}} p_0$.

The stable stationary solution corresponding to the 2f synchronized state has the phase difference given by $\psi_s = \arccos\left(\dfrac{\Delta\omega}{\Omega}\right)$ with $\Omega = \dfrac{2N p_0}{\Gamma_p} \mathcal{F}$ and a stationary shift in power given by $\delta p_s = \dfrac{\Delta\omega}{2N}$.